\documentclass[aps,prb,twocolumn,groupedaddress,showpacs,showkeys]{revtex4}

\usepackage{graphicx}  
\usepackage{dcolumn}   
\usepackage{bm}        
\usepackage{multirow}  
\usepackage{color}     
\usepackage{rotating}  
\usepackage{amsmath}
\usepackage{amssymb}
\usepackage{relsize}

\newcommand{\be}{\begin{equation}}
\newcommand{\ee}{\end{equation}}

\newcommand{\beq}{\begin{eqnarray}}
\newcommand{\eeq}{\end{eqnarray}}
\newcommand{\ba}{\begin{array}}
\newcommand{\ea}{\end{array}}

\def\H1{\widehat{H}_1}

\newcommand{\e}{\ensuremath{\mathrm{e}}}
\renewcommand{\d}{\ensuremath{\mathrm{d}}}

\begin{document}

\title{Coulomb drag between massless and massive fermions.}
\author{Benedikt Scharf and Alex Matos-Abiague}
\affiliation{Institute for Theoretical Physics, University of Regensburg, 93040 Regensburg, Germany}
\date{\today}

\begin{abstract}
We theoretically investigate the frictional drag induced by the Coulomb interaction between spatially separated massless and massive fermions in the Boltzmann regime and at low temperatures. As a model system, we use a double-layer structure composed of a two-dimensional electron gas (2DEG) and a $n$-doped graphene layer. We analyze this system numerically and also present analytical formulae for the drag resistivity in the limit of large and small interlayer separation. Both, the temperature and density dependence are investigated and compared to 2DEG-2DEG and graphene-graphene double-layer structures. Whereas the density dependence of the transresistivity for small interlayer separation differs already in the leading order for each of those three structures, we find the leading order contribution of the density dependence in the large interlayer separation limit to exhibit the same density dependence in each case. In order to distinguish between the different systems in the large interlayer separation limit, we also investigate the subleading contribution to the transresistivity. Furthermore, we study the Coulomb drag in a double-layer structure consisting of $n$-doped bilayer and monolayer graphene, which we find to possess the same qualitative behavior as the 2DEG-graphene system.
\end{abstract}


\maketitle

\section{Introduction}\label{Sec:Intro}
The transport properties of double-layer systems, in which carriers are confined to nearby parallel planes, have received considerable attention since the earlier proposal by Pogrebinski\u{\i}\cite{Pogrebinskii1977:SPS} of employing a bilayer system for measuring the frictional drag. Drag measurements are performed by driving a current $\mathbf{j}_a$ through one of the layers (the active layer) and measuring the electric field $\mathbf{E}_p$ induced in the other layer (the passive layer) due to interlayer momentum transfer. The drag transresistivity (also called drag coefficient, or simply, drag) is defined as $\rho_{D} = E_{p}/j_{a}$. The measurement of the frictional drag can provide valuable information about the density and temperature dependence of the carrier-carrier interaction in 2D systems. In particular, the frictional drag due to the interlayer carrier-carrier Coulomb interaction in double-layer semiconductor systems has been investigated in great detail.\cite{Gramila1991:PRL,Zheng1993:PRB,Jauho1993:PRB,Kamenev1995:PRB,Flensberg1995:PRB,Rojo1999:JPCM,Pillarisetty2002:PRL,Hwang2003:PRL,Badalyan2007:PRB}

With the recent progress in the physics of graphene, much attention has been devoted to the investigation and understanding of the frictional drag in spatially separated double-layer graphene systems\cite{Narozhny2007:PRB,Tse2007:PRB,Kim2011:PRB,Katsnelson2011:PRB,Peres2011:EPL,Hwang2011:PRB,Narozhny2012:PRB,Carrega2012:arXiv,Amorim2012:arXiv} as well as in structures comprising two bilayer graphene (BLG) sheets isolated from each other by a spacer.\cite{Hwang2011:PRB} Moreover, in the limit of low temperatures and large interlayer distances, a generic formula has been derived for the leading order of the asymptotic behavior in the limit of large interlayer separation for systems where each layer $l$ is described by an energy dispersion of the form $\epsilon_{\mathbf{k}}^{\mathsmaller{l}}\propto k^{2-\xi_\mathsmaller{l}}$ ($\xi_\mathsmaller{l}$ is a layer specific constant) and a momentum-dependent relaxation time $\tau_{\mathsmaller{l}}\left(\mathbf{k}\right)$.\cite{Amorim2012:arXiv}

There is a fundamental difference between the carriers in graphene and those in a two-dimensional electron gas (2DEG) or in BLG. While in graphene the carriers can be interpreted as massless fermions with a linear dispersion, in the 2DEG and BLG the carriers exhibit a parabolic dispersion and have a finite effective mass. Thus, most of the previously reported investigations of the frictional drag have been limited to the case of interaction between massive fermions (in the case of 2DEG-2DEG or BLG-BLG double-layer systems) or between massless fermions (in the case of graphene-graphene double-layer structures). In what follows, we will refer to the former and later cases as massive-massive and massless-massless systems, respectively. By assembling a double-layer structure consisting of a graphene layer and a 2DEG layer, it might also be possible to create a setup where the carrier densities are significantly different in both layers, a case difficult to achieve if both layers consist of the same material.

In the present paper, we investigate massless-massive systems in which the frictional drag is induced by the Coulomb interaction between massless and massive fermions, a case that till now has remained largely unexplored. Here, we restrict ourselves to the discussion of low temperatures and the case where both layers are in the ballistic/Boltzmann regime. As a prototype system, we consider first a double-layer structure consisting of a 2DEG formed in a GaAs quantum well and a closely located $n$-doped graphene layer. We then compute the transresistivity for such a system and investigate its dependence on temperature and carrier concentrations. We also provide analytical formulae describing the asymptotic behavior of the transresistivity in the large and small interlayer separation limits and compare our results with those corresponding to 2DEG-2DEG, graphene-graphene, and BLG-BLG double-layer structures. We show that in the small interlayer separation limit, already at the leading order the transresistivity scales with the carrier densities differently for all the three massive-massive, massless-massless, and massless-massive systems. However, in the large interlayer separation limit the three kinds of systems exhibit the same asymptotic behavior in the leading order and differences appear only when the subleading correction is taken into account. As an alternative to the 2DEG-graphene structure, we also investigate the drag transresistivity in a double-layer structure consisting of $n$-doped bilayer and monolayer graphene isolated from each other by a spacer. Such a massless-massive system exhibits the same qualitative behavior as the 2DEG-graphene structure.

The manuscript is organized as follows: In Sec.~\ref{Sec:2DEG_graphene}, following the introduction of the model and the theoretical framework, the Coulomb drag in 2DEG/(monolayer) graphene systems is discussed. This discussion is extended to a bilayer graphene/(monolayer) graphene system in Sec.~\ref{Sec:bilayer_monolayer_graphene}. Corrections to the asymptotic behavior in the limit of large interlayer distances are considered in Sec.~\ref{Sec:Corrections}. A short summary concludes the manuscript.

\section{Drag resistivity in 2DEG/monolayer graphene systems}\label{Sec:2DEG_graphene}

\begin{figure}[t]
\centering
\includegraphics*[width=7cm]{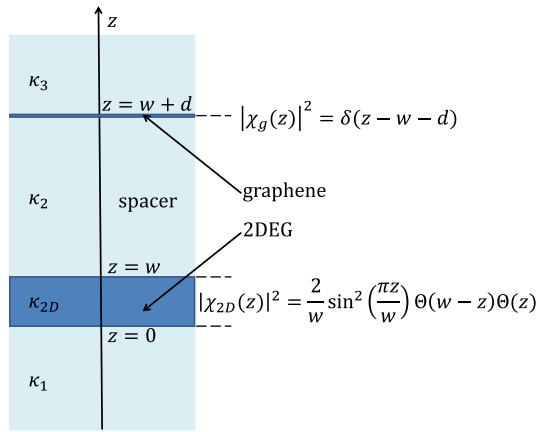}
\caption{(Color online) Schematic illustration of the geometry considered. The 2DEG is located within a quantum well of width $w$ and its localization along the $z$ direction is described by $\chi_{\mathsmaller{\mathrm{2D}}}(z)$, whereas the location of the graphene sheet is given by $\chi_g(z)$. The relative dielectric constants of the structure are given by $\kappa_1$, $\kappa_2$, $\kappa_3$, and $\kappa_\mathsmaller{\mathrm{2D}}$.}\label{fig:Graphene_vs_2DEG}
\end{figure}

\subsection{Model}
In this Section, we investigate a double-layer structure consisting of a 2DEG, located within a quantum well of width $w$, and one $n$-doped layer of graphene. Both electronic systems are separated by a spacer of width $d$ and embedded in a larger structure. The relative dielectric constants in the different regions of the structure are denoted by $\kappa_1$, $\kappa_\mathsmaller{\mathrm{2D}}$, $\kappa_2$, and $\kappa_3$  (see Fig.~\ref{fig:Graphene_vs_2DEG}).

Throughout this manuscript, we consider the case where both layers are within the Boltzmann regime (that is, the regime in which the Fermi wavevector is much larger than the inverse mean free path) and have the same temperature $T$. Furthermore, these temperatures are assumed to be low, that is,
\begin{equation}\label{small_T}
k_\mathsmaller{\mathrm{B}}T\ll\epsilon_{\mathsmaller{\mathrm{F}}}^\mathsmaller{\mathrm{2D/g}},
\end{equation}
where $k_\mathsmaller{\mathrm{B}}$ and $\epsilon_{\mathsmaller{\mathrm{F}}}^\mathsmaller{\mathrm{2D/g}}$ denote the Boltzmann constant and the Fermi energies of the 2DEG and graphene layers, respectively. In what follows, graphene is assumed to be the active layer, while the 2DEG is taken as the passive one.

A peculiar property of the considered double-layer structure is the presence of interactions between two kinds of carriers, massive and massless fermions. Indeed, the carriers in the 2DEG are massive fermions with effective mass $m^{\ast}$ and dispersion relation
\begin{equation}\label{dispersion-2deg}
    \epsilon_{\mathbf{k}}^{\mathsmaller{\mathrm{2D}}}=\frac{\hbar^{2}k^{2}}{2m^{\ast}},
\end{equation}
while the carriers in the graphene layer are massless fermions with Fermi velocity $v_{\mathsmaller{\mathrm{F}}}^{\mathsmaller{\mathrm{g}}}\approx 10^{6}\textrm{m/s}$ and the linear dispersion
\begin{equation}\label{dispersion-g}
    \epsilon_{\mathbf{k}}^{\mathsmaller{\mathrm{g}}}=\hbar v_{\mathsmaller{\mathrm{F}}}^{\mathsmaller{\mathrm{g}}} |\mathbf{k}|.
\end{equation}

For most practical situations, the interlayer distance ($d$) is such that the interlayer Coulomb interaction is weak. Thus, a lowest-order perturbation theory in the interlayer potential suffices and the transresistivity $\rho^{ij}_\mathsmaller{\mathrm{D}}$ is found to be given by,\cite{Kamenev1995:PRB,Flensberg1995:PRB}
\begin{equation}\label{transresistivity_LRT}
\begin{aligned}
\rho^{ij}_\mathsmaller{\mathrm{D}}=&\frac{-1}{16\pi Sk_{\mathsmaller{B}}T\sigma_{\mathsmaller{\mathrm{2D}}}\sigma_\mathsmaller{\mathrm{g}}}\\
&\times\sum\limits_{\mathbf{q}}\int\limits_{-\infty}^{\infty}\d\omega\frac{\Gamma_{\mathsmaller{\mathrm{2D}}}^{i}\left(\mathbf{q},\omega\right)\Gamma_{\mathsmaller{\mathrm{g}}}^{j}\left(\mathbf{q},\omega\right)\left|U_{\mathsmaller{\mathrm{2Dg}}}\left(\mathbf{q},\omega\right)\right|^2}{\sinh^2\left(\hbar\omega/2k_\mathsmaller{\mathrm{B}}T\right)},
\end{aligned}
\end{equation}
where $S$ is the cross section area of the layers, $U_{\mathsmaller{\mathrm{2Dg}}}\left(\mathbf{q},\omega\right)$ is the screened interlayer potential between the 2DEG and graphene layers, and $\sigma_\mathsmaller{\mathrm{2D/g}}$ and $\Gamma_{\mathsmaller{\mathrm{2D/g}}}^{i}\left(\mathbf{q},\omega\right)$ denote the Drude conductivity and the $i$-th component of the nonlinear susceptibility in the 2DEG and graphene layers, respectively. The Drude conductivities are given by $\sigma_{\mathsmaller{\mathrm{2D/g}}}=e^2\epsilon_{\mathsmaller{\mathrm{F}}}^{\mathsmaller{\mathrm{2D/g}}}\tau_{\mathsmaller{\mathrm{2D/g}}}/(\pi\hbar^2)$, where the momentum relaxation times (at the Fermi energy), $\tau_\mathsmaller{\mathrm{2D/g}}$, are defined below. Furthermore, we assume that there is no electron tunneling between both layers, so the Fermi energies $\epsilon_{\mathsmaller{\mathrm{F}}}^\mathsmaller{\mathrm{2D/g}}$ can be set independently from each other in each layer. Due to the factor $1/\sinh^2\left(\hbar\omega/2k_\mathsmaller{\mathrm{B}}T\right)$, only small values of $\omega$ contribute to the transresistivity at low temperatures, while the screened interlayer potential (see below) restricts the momentum integration to small values of $q$. Therefore, we can approximate the nonlinear susceptibilities by their respective expressions in the limit of low energies and long wavelengths.

\subsection{Nonlinear susceptibilities}
Before we continue with those expressions, we briefly mention the general expression for the nonlinear susceptibility within the Boltzman limit, that is, the regime of $k_\mathsmaller{\mathrm{F}}l\gg1$ or $\omega\tau\gg1$ with $\tau$, $l=v_\mathsmaller{\mathrm{F}}\tau$, $k_\mathsmaller{\mathrm{F}}$, and $v_\mathsmaller{\mathrm{F}}$ being the scattering time at the Fermi level, the mean free path, and the Fermi wavevector and velocity, respectively (for brevity, we suppress the index label denoting the system here and in the following). In this limit, the nonlinear susceptibilities of monolayer and bilayer graphene as well as 2DEGs can be written as
\begin{equation}\label{nonlinear_susceptibility_general}
\begin{aligned}
\bm{\Gamma}&\left(\mathbf{q},\omega\right)=\\
&-\frac{2eg_sg_v}{S\hbar}\sum\limits_{\mathbf{k},\lambda,\lambda'}\mathrm{Im}\left[\frac{\lambda\left(f^{\mathsmaller{\lambda}}_{\mathsmaller{\mathbf{k}}}-f^{\mathsmaller{\lambda'}}_{\mathsmaller{\mathbf{k}+\mathbf{q}}}\right)\bm{\mu}^{\mathsmaller{\lambda,\lambda'}}_{\mathsmaller{\mathbf{k},\mathbf{k}+\mathbf{q}}}F^{\mathsmaller{\lambda,\lambda'}}_{\mathsmaller{\mathbf{k},\mathbf{k}+\mathbf{q}}}}{\hbar\omega+\epsilon_{\lambda,\mathbf{k}}-\epsilon_{\lambda',\mathbf{k}+\mathbf{q}}+\mathrm{i}0^+}\right],
\end{aligned}
\end{equation}
where $e=|e|$ denotes the absolute value of the electron charge, $\lambda$ and $\lambda'$ band labels, $\epsilon_{\lambda,\mathbf{k}}=\lambda\epsilon_\mathbf{k}$ the energy in a given band with $\epsilon_\mathbf{k}$ being the dispersion of the system investigated [that is, Eq.~(\ref{dispersion-2deg}) for 2DEGs and bilayer graphene and Eq.~(\ref{dispersion-g}) for monolayer graphene], and $f^{\mathsmaller{\lambda}}_{\mathsmaller{\mathbf{k}}}$ the Fermi-Dirac distribution function for the energy $\epsilon_{\lambda,\mathbf{k}}$. For monolayer and bilayer graphene, $\lambda=\pm1$ describes the valence and conduction bands, while for 2DEGs $\lambda=1$. The spin degeneracy is described by the factor $g_s=2$ for 2DEGs as well as monolayer and bilayer graphene, whereas $g_v$ describes the valley degeneracy factor, which is $g_v=1$ in 2DEGs and $g_v=2$ in monolayer and bilayer graphene. The factor $F^{\mathsmaller{\lambda,\lambda'}}_{\mathsmaller{\mathbf{k},\mathbf{k}+\mathbf{q}}}$, which arises due to the overlap of the wavefunctions, is unity for 2DEGs, but
\begin{equation}\label{wf-overlap-g}
F^{\mathsmaller{\lambda,\lambda'}}_{\mathsmaller{\mathbf{k},\mathbf{k}+\mathbf{q}}}=\frac{1+\lambda\lambda'\cos\left(\Theta_\mathsmaller{\mathbf{k}+\mathbf{q}}-\Theta_\mathsmaller{\mathbf{k}}\right)}{2}
\end{equation}
and
\begin{equation}\label{wf-overlap-bg}
F^{\mathsmaller{\lambda,\lambda'}}_{\mathsmaller{\mathbf{k},\mathbf{k}+\mathbf{q}}}=\frac{1+\lambda\lambda'}{2}-\frac{\lambda\lambda'q^2\sin^2\left(\Theta_\mathsmaller{\mathbf{k}}-\Theta_\mathsmaller{\mathbf{q}}\right)}{|\mathbf{k}+\mathbf{q}|^2}
\end{equation}
for monolayer and bilayer graphene, respectively.\cite{Hwang2011:PRB} Here, $\Theta_\mathsmaller{\mathbf{k}}$ is the azimuthal angle of $\mathbf{k}$ in momentum space. We have also introduced the quantity $\bm{\mu}^{\mathsmaller{\lambda,\lambda'}}_{\mathsmaller{\mathbf{k},\mathbf{k}+\mathbf{q}}}$, which reads as
\begin{equation}\label{current-tau-2DEG-bg}
\bm{\mu}^{\mathsmaller{\lambda,\lambda'}}_{\mathsmaller{\mathbf{k},\mathbf{k}+\mathbf{q}}}=\frac{\hbar^2}{m}\left[\tau(\mathbf{k})\mathbf{k}-\lambda\lambda'\tau(\mathbf{k}+\mathbf{q})\left(\mathbf{k}+\mathbf{q}\right)\right]
\end{equation}
for 2DEGs and bilayer graphene and
\begin{equation}\label{current-tau-g}
\bm{\mu}^{\mathsmaller{\lambda,\lambda'}}_{\mathsmaller{\mathbf{k},\mathbf{k}+\mathbf{q}}}=\hbar v_\mathsmaller{\mathrm{F}}\left[\tau(\mathbf{k})\frac{\mathbf{k}}{|\mathbf{k}|}-\lambda\lambda'\tau(\mathbf{k}+\mathbf{q})\frac{\mathbf{k}+\mathbf{q}}{|\mathbf{k}+\mathbf{q}|}\right]
\end{equation}
for monolayer graphene. Equations~(\ref{current-tau-2DEG-bg}) and~(\ref{current-tau-g}) contain the scattering time $\tau(\mathbf{k})$, which in general can be momentum-dependent. Since we are interested in the limit of low temperatures, we will use the expressions obtained from Eq.~(\ref{nonlinear_susceptibility_general}) for $T\to0$ and low energies and long wavelengths for the rest of the manuscript (from here on, we also restore the index label denoting the system).

In 2DEGs, the main effects of both short range and screened Coulomb impurities can be properly described by considering the relaxation time $\tau_\mathsmaller{\mathrm{2D}}$ to be momentum independent. Here, $\tau_\mathsmaller{\mathrm{2D}}$ denotes the relaxation time at the Fermi level---a consequence of the fact that at low temperatures the nonlinear susceptibility is determined by electrons at the Fermi surface. In this case, the nonlinear susceptibility in the limit of low energies and long wavelengths reads as\cite{Kamenev1995:PRB,Flensberg1995:PRB,Giuliani:2005}
\begin{equation}\label{nonlinear_susceptibility_2DEG}
\Gamma_{\mathsmaller{\mathrm{2D}}}^{i}\left(\mathbf{q},\omega\right)=-\frac{2e\omega\tau_\mathsmaller{\mathrm{2D}}}{\hbar\pi v_{\mathsmaller{\mathrm{F}}}^{\mathsmaller{\mathrm{2D}}}}\frac{q_i}{q}\Lambda_{\mathsmaller{\mathrm{2D}}}(q),
\end{equation}
where
\begin{equation}\label{g_2DEG}
\Lambda_{\mathsmaller{\mathrm{2D}}}(q)=\frac{\Theta\left(2k_{\mathsmaller{\mathrm{F}}}^\mathsmaller{\mathrm{2D}}-q\right)}{\sqrt{1-\left(\frac{q}{2k_{\mathsmaller{\mathrm{F}}}^\mathsmaller{\mathrm{2D}}}\right)^2}},
\end{equation}
and $v_{\mathsmaller{\mathrm{F}}}^\mathsmaller{\mathrm{2D}}$ and $k_{\mathsmaller{\mathrm{F}}}^\mathsmaller{\mathrm{2D}}$ are the Fermi velocity and wavevector in the 2DEG layer.

Contrary to the case of the 2DEG, the relaxation time describing electron-impurity scattering in graphene, which is proportional to the momentum, that is, $\tau_{\mathsmaller{\mathrm{g}}}\left(\mathbf{k}\right)=\tau_0k$, with $\tau_0$ being a constant of proportionality,\cite{Novoselov2005:Nature,Tan2007:PRL,Chen2008:NatPhys,Peres2010:RMP,DasSarma2011:RMP,Peres2011:EPL} is widely used as a model for the relaxation time in graphene. In this case, the nonlinear susceptibility in the limit of low energies can be written as\cite{Narozhny2012:PRB,Carrega2012:arXiv,Amorim2012:arXiv}
\begin{equation}\label{nonlinear_susceptibility_graphene}
\Gamma_{\mathsmaller{\mathrm{g}}}^{i}\left(\mathbf{q},\omega\right)=-\frac{4e\omega\tau_g}{\hbar\pi v_{\mathsmaller{\mathrm{F}}}^\mathsmaller{\mathrm{g}}}\frac{q_i}{q}\Lambda_{\mathsmaller{\mathrm{g}}}(q),
\end{equation}
where
\begin{equation}\label{g_graphene}
\Lambda_{\mathsmaller{\mathrm{g}}}(q)=\sqrt{1-\left(\frac{q}{2k_{\mathsmaller{\mathrm{F}}}^\mathsmaller{\mathrm{g}}}\right)^2}\Theta\left(2k_{\mathsmaller{\mathrm{F}}}^\mathsmaller{\mathrm{g}}-q\right).
\end{equation}
Here, $v_{\mathsmaller{\mathrm{F}}}^\mathsmaller{\mathrm{g}}$ and $k_{\mathsmaller{\mathrm{F}}}^\mathsmaller{\mathrm{g}}$ denote the Fermi velocity and wavevector in graphene and the relaxation time at the Fermi level is given by $\tau_\mathsmaller{\mathrm{g}}=\tau_0k_{\mathsmaller{\mathrm{F}}}^\mathsmaller{\mathrm{g}}$. We note that Eqs.~(\ref{nonlinear_susceptibility_graphene}) and~(\ref{g_graphene}) are the same results one would have obtained if the relaxation time in graphene had been assumed as constant (that is, $\tau_{\mathsmaller{\mathrm{g}}}\left(\mathbf{k}\right)=\tau_{\mathsmaller{\mathrm{g}}}=\mathrm{const.}$).\cite{Tse2007:PRB} Indeed, it has been shown in Refs.~\onlinecite{Narozhny2012:PRB,Carrega2012:arXiv,Amorim2012:arXiv} that---if isotropic relaxation times are assumed---the form of $\tau_{\mathsmaller{\mathrm{g}}}\left(|\mathbf{k}|\right)$ as a function of the momentum does not affect the low-temperature limit of the nonlinear susceptibility, Eq.~(\ref{nonlinear_susceptibility_graphene}), and one can replace the momentum-dependent relaxation time by its value at the Fermi level. Moreover, one can notice that, within the limit of Eqs.~(\ref{nonlinear_susceptibility_2DEG}) and~(\ref{nonlinear_susceptibility_graphene}), the momentum integration is cut off for $q>2k_{\mathsmaller{\mathrm{F}}}^\mathsmaller{\mathrm{2D/g}}$.

\subsection{Interlayer potential and transresistivity}
The screened interlayer potential can be found by solving the corresponding Dyson equation and can be written as
\begin{equation}\label{screened_interlayer_potential}
U_{\mathsmaller{\mathrm{2Dg}}}\left(\mathbf{q},\omega\right)=\frac{U_{\mathsmaller{\mathrm{2Dg}}}^{(0)}\left(\mathbf{q}\right)}{\epsilon_{\mathsmaller{\mathrm{2Dg}}}\left(\mathbf{q},\omega\right)}
\end{equation}
with
\begin{equation}\label{interlayer_potential_dielectric}
\begin{aligned}
\epsilon_{\mathsmaller{\mathrm{2Dg}}}\left(\mathbf{q},\omega\right)=&\left[1+U_{\mathsmaller{\mathrm{2D}}}^{(0)}\left(\mathbf{q}\right)\Pi_{\mathsmaller{\mathrm{2D}}}\left(\mathbf{q},\omega\right)\right]\\
&\times\left[1+U_\mathsmaller{\mathrm{g}}^{(0)}\left(\mathbf{q}\right)\Pi_g\left(\mathbf{q},\omega\right)\right]\\
&-\left|U_{\mathsmaller{\mathrm{2Dg}}}^{(0)}\left(\mathbf{q}\right)\right|^2\Pi_{\mathsmaller{\mathrm{2D}}}\left(\mathbf{q},\omega\right)\Pi_g\left(\mathbf{q},\omega\right),
\end{aligned}
\end{equation}
where $\Pi_{\mathsmaller{\mathrm{2D/g}}}\left(\mathbf{q},\omega\right)$ are the polarization functions of the individual layers, for each of which we use the respective expressions obtained from the random phase approximation (RPA) at zero temperature. The bare intralayer and interlayer Coulomb potentials can be written as $U_{\mathsmaller{\mathrm{2D/g}}}^{(0)}\left(\mathbf{q}\right)=(4\pi e^2/q)f_{\mathsmaller{\mathrm{2D/g}}}\left(qd,qw\right)$ and $U_{\mathsmaller{\mathrm{2Dg}}}^{(0)}\left(\mathbf{q}\right)=(8\pi e^2/q)f_{\mathsmaller{\mathrm{2Dg}}}\left(qd,qw\right)$, respectively, where the form factors $f_{\mathsmaller{\mathrm{2D/g}}}\left(qd,qw\right)$ and $f_{\mathsmaller{\mathrm{2Dg}}}\left(qd,qw\right)$ are determined by solving the Poisson equation of the system (see Appendix~\ref{Sec:AppendixCoulomb}). Since at low temperatures only small values of $\omega$ contribute to Eq.~(\ref{transresistivity_LRT}), we approximate the dynamic by the static polarization functions, that is, we replace $U_{\mathsmaller{\mathrm{2Dg}}}\left(\mathbf{q},\omega\right)$ by the static interlayer potential $U_{\mathsmaller{\mathrm{2Dg}}}\left(\mathbf{q},0\right)$.

\begin{table*}
\begin{center}
\begin{tabular}{|c||c|c|c|}
\hline
system & $\rho_{\mathsmaller{D}}$ & $\rho_{\mathsmaller{D}}$ & $\rho_{\mathsmaller{D}}$\\
(active-passive) & ($q_{\mathsmaller{\mathrm{TF}}}^{\mathsmaller{\mathrm{a/p}}}d\ll 1$, $k_{\mathsmaller{\mathrm{F}}}^{\mathsmaller{\mathrm{a/p}}}d\ll 1$) & ($q_{\mathsmaller{\mathrm{TF}}}^{\mathsmaller{\mathrm{a/p}}}d\ll 1$, $k_{\mathsmaller{\mathrm{F}}}^{\mathsmaller{\mathrm{a/p}}}d\gg 1$) & ($q_{\mathsmaller{\mathrm{TF}}}^{\mathsmaller{\mathrm{a/p}}}d\gg 1$, $k_{\mathsmaller{\mathrm{F}}}^{\mathsmaller{\mathrm{a/p}}}d\gg 1$)\\
\hline\hline
massive-massive & $\propto\frac{1}{n_{\mathsmaller{\mathrm{a}}}^3}$ & $\propto\frac{\ln[(q_{\mathsmaller{\mathrm{TF}}}^{\mathsmaller{\mathrm{a}}}+q_{\mathsmaller{\mathrm{TF}}}^{\mathsmaller{\mathrm{p}}})d]}{(n_{_a}n_{_p})^{3/2}}$ & $\propto1/[(n_{_a}n_{_p})^{3/2}d^{4}]$\\
\hline
massless-massless & $\propto\frac{1}{n_\mathsmaller{\mathrm{a}}}$ & $\propto\frac{\ln[(\sqrt{n_{_a}}+\sqrt{n_{_p}})d/\overline{\kappa}]}{\sqrt{n_{_a}n_{_p}}}$ & $\propto1/[(n_{_a}n_{_p})^{3/2}d^{4}]$\\
\hline
massless-massive & $\propto\frac{1}{n_{\mathsmaller{\mathrm{a}}}^2}$ & $\propto\frac{\ln[(\alpha\sqrt{\pi n_{_a}}+q_{\mathsmaller{\mathrm{TF}}}^{\mathsmaller{\mathrm{p}}})d]}{\sqrt{n_{_a}n_{_p}^{3}}}$ & $\propto1/[(n_{_a}n_{_p})^{3/2}d^{4}]$\\
\hline
\end{tabular}
\end{center}
\caption{Asymptotic behavior of the transresistivity $\rho_{\mathsmaller{D}}$ as a function of the densities and interlayer distance for different systems. In the limit $q_{\mathsmaller{\mathrm{TF}}}^{\mathsmaller{\mathrm{a/p}}}d\gg 1$, the three systems exhibit identical behavior in the leading order and one needs to consider the subleading correction $\Delta\rho_{\mathsmaller{D}}$ in order to see differences in the transresistivity (see Sec.~\ref{Sec:Corrections}). The subscript $a$ ($p$) refers to the active (passive) layer. The screened Thomas-Fermi wavevectors, particle densities, interlayer distance, and average dielectric constant are denoted, respectively, by $q_{\mathsmaller{\mathrm{TF}}}^{\mathsmaller{\mathrm{a/p}}}$, $n_{_{a/p}}$, $d$, and $\overline{\kappa}$. We have also introduced the constant $\alpha=4e^{2}/(\overline{\kappa}\hbar v_{\mathsmaller{\mathrm{F}}}^{0})$ with $v_{\mathsmaller{\mathrm{F}}}^{0}$ denoting the Fermi velocity of the massless particles. Since there is no general analytical formula for the small interlayer separation limit, $q_{\mathsmaller{\mathrm{TF}}}^{\mathsmaller{\mathrm{a/p}}}d\ll 1$ and $k_{\mathsmaller{\mathrm{F}}}^{\mathsmaller{\mathrm{a/p}}}d\ll 1$ (first column), we provide formulae for the case of high densities and $k_{\mathsmaller{\mathrm{F}}}^{\mathsmaller{\mathrm{a}}}=k_{\mathsmaller{\mathrm{F}}}^{\mathsmaller{\mathrm{p}}}$, that is, $n_\mathsmaller{\mathrm{a}}=2n_\mathsmaller{\mathrm{p}}$ for a 2DEG-graphene system and $n_\mathsmaller{\mathrm{a}}=n_\mathsmaller{\mathrm{p}}$ for a BLG-graphene system (see Appendix~\ref{Sec:AppendixLimitingCases}).}
\label{tab:comparison}
\end{table*}

Within the above approximations, the transresistivity is diagonal because the nonlinear susceptibilities and the screened interlayer potential are isotropic, and it is therefore enough to calculate $\rho_\mathsmaller{\mathrm{D}}=\rho^{xx}_\mathsmaller{\mathrm{D}}$. Thus, we obtain the transresistivity at low temperatures,
\begin{equation}\label{transresistivity_final}
\rho_\mathsmaller{\mathrm{D}}=-\frac{h}{e^2}\frac{4\pi}{3}\frac{\left(k_{\mathsmaller{B}}T\right)^2F\left(Q_{\mathsmaller{\mathrm{TF}}}^{\mathsmaller{\mathrm{2D}}}d,Q_{\mathsmaller{\mathrm{TF}}}^{\mathsmaller{\mathrm{g}}}d,w/d\right)}{\epsilon_{\mathsmaller{\mathrm{F}}}^{\mathsmaller{\mathrm{2D}}}\epsilon_{\mathsmaller{\mathrm{F}}}^{\mathsmaller{\mathrm{g}}}\left(k_{\mathsmaller{\mathrm{F}}}^{\mathsmaller{\mathrm{2D}}}d\right)\left(k_{\mathsmaller{\mathrm{F}}}^{\mathsmaller{\mathrm{g}}}d\right)\left(Q_{\mathsmaller{\mathrm{TF}}}^{\mathsmaller{\mathrm{2D}}}d\right)\left(Q_{\mathsmaller{\mathrm{TF}}}^{\mathsmaller{\mathrm{g}}}d\right)},
\end{equation}
where $Q_{\mathsmaller{\mathrm{TF}}}^{\mathsmaller{\mathrm{2D/g}}}=2\pi e^2\nu_\mathsmaller{\mathrm{2D/g}}$ and $\nu_\mathsmaller{\mathrm{2D/g}}$ denote the bare Thomas-Fermi wavevector and the (total) density of states at the Fermi level in each individual layer. The function $F(y_{\mathsmaller{\mathrm{2D}}},y_\mathsmaller{\mathrm{g}},r)$ is given by the integral
\begin{widetext}
\begin{equation}\label{transresistivity_integral}
F(y_{\mathsmaller{\mathrm{2D}}},y_\mathsmaller{\mathrm{g}},r)=\int\limits_{0}^{\infty}\d x\frac{\left[y_{\mathsmaller{\mathrm{2D}}}y_\mathsmaller{\mathrm{g}}f_{\mathsmaller{\mathrm{2Dg}}}(x,xr)\right]^2\,x^3\,\Lambda_{\mathsmaller{\mathrm{2D}}}(x/d)\Lambda_{\mathsmaller{\mathrm{g}}}(x/d)}{\left\{\left[x+2y_{\mathsmaller{\mathrm{2D}}}f_{\mathsmaller{\mathrm{2D}}}(x,xr)\tilde{\Pi}_{\mathsmaller{\mathrm{2D}}}(x)\right]\left[x+2y_gf_g(x,xr)\tilde{\Pi}_g(x)\right]-16y_{\mathsmaller{\mathrm{2D}}}y_g{f^2_{\mathsmaller{\mathrm{2Dg}}}}(x,xr)\tilde{\Pi}_{\mathsmaller{\mathrm{2D}}}(x)\tilde{\Pi}_g(x)\right\}^2},
\end{equation}
\end{widetext}
where $y_{\mathsmaller{\mathrm{2D/g}}}=Q_{\mathsmaller{\mathrm{TF}}}^{\mathsmaller{\mathrm{2D/g}}}d$, $r=w/d$, $x=qd$, and
\begin{equation}\label{polarization_function_2DEG}
\begin{aligned}
\tilde{\Pi}_{\mathsmaller{\mathrm{2D}}}(x)&=\frac{\Pi_{\mathsmaller{\mathrm{2D}}}(x/d,0)}{\nu_\mathsmaller{\mathrm{2D}}}\\
&=1-\Theta\left(\frac{x}{2k_{\mathsmaller{\mathrm{F}}}^{\mathsmaller{\mathrm{2D}}}d}-1\right)\sqrt{1-\left(\frac{2k_{\mathsmaller{\mathrm{F}}}^{\mathsmaller{\mathrm{2D}}}d}{x}\right)^2}
\end{aligned}
\end{equation}
and
\begin{equation}\label{polarization_function_graphene}
\begin{aligned}
\tilde{\Pi}_\mathsmaller{\mathrm{g}}(x)&=\frac{\Pi_\mathsmaller{\mathrm{g}}(x/d,0)}{\nu_\mathsmaller{\mathrm{g}}}\\
&=1+\Theta\left(\frac{x}{2k_{\mathsmaller{\mathrm{F}}}^{\mathsmaller{\mathrm{g}}}d}-1\right)\frac{x}{4k_{\mathsmaller{\mathrm{F}}}^{\mathsmaller{\mathrm{g}}}d}\\
&\times\left[\arccos\left(\frac{2k_{\mathsmaller{\mathrm{F}}}^{\mathsmaller{\mathrm{g}}}d}{x}\right)-\frac{2k_{\mathsmaller{\mathrm{F}}}^{\mathsmaller{\mathrm{g}}}d}{x}\sqrt{1-\left(\frac{2k_{\mathsmaller{\mathrm{F}}}^{\mathsmaller{\mathrm{g}}}d}{x}\right)^2}\right]
\end{aligned}
\end{equation}
are the static, dimensionless polarization functions of the 2DEG\cite{Giuliani:2005,Stern1967:PRL} and graphene,\cite{Wunsch2006:NJoP,Hwang2007:PRB} respectively.

\subsection{Asymptotic behavior}\label{Sec:Asymptotics}
In general, the integration in Eq.~(\ref{transresistivity_integral}), and therefore the transresistivity in Eq.~(\ref{transresistivity_final}), have to be computed numerically. However, simplified analytical expressions describing the asymptotic behavior of the drag resistivity as described by Eqs.~(\ref{transresistivity_final}) and~(\ref{transresistivity_integral}) can be obtained in the limits of large and small interlayer distances by replacing each of the different relative dielectric constants $\kappa_1$, $\kappa_2$, $\kappa_3$, and $\kappa_\mathsmaller{\mathrm{2D}}$ by an average relative dielectric constant of the entire structure, $\overline{\kappa}$. Then, we can introduce the screened Thomas-Fermi wavevectors, $q_{\mathsmaller{\mathrm{TF}}}^{\mathsmaller{\mathrm{2D/g}}}=Q_{\mathsmaller{\mathrm{TF}}}^{\mathsmaller{\mathrm{2D/g}}}/\overline{\kappa}$. Moreover, the form factors reduce to
\begin{equation}\label{form_factor_2D}
\begin{aligned}
f_{\mathsmaller{\mathrm{2D}}}=\frac{1}{2\overline{\kappa}}\frac{32\pi^4\left(\e^{-xr}-1+xr\right)+20\pi^2\left(xr\right)^3+3\left(xr\right)^5}{\left(xr\right)^2\left[4\pi^2+\left(xr\right)^2\right]^2},
\end{aligned}
\end{equation}
\begin{equation}\label{form_factor_g}
f_g=\frac{1}{2\overline{\kappa}},
\end{equation}
and
\begin{equation}\label{form_factor_interlayer}
\begin{aligned}
f_{\mathsmaller{\mathrm{2Dg}}}=\frac{\e^{-x}}{4\overline{\kappa}}\frac{4\pi^2\left(1-\e^{-xr}\right)}{xr\left[4\pi^2+\left(xr\right)^2\right]}.
\end{aligned}
\end{equation}
Since the upper boundary of the integral given by Eq.~(\ref{transresistivity_integral}) is restricted by the minimum of the Fermi wavevectors [$\mathrm{min}\left(2k_{\mathsmaller{\mathrm{F}}}^{\mathsmaller{\mathrm{2D}}}d,2k_{\mathsmaller{\mathrm{F}}}^{\mathsmaller{\mathrm{g}}}d\right)$], the polarization functions can be replaced by their long wavelength limits, that is, $\tilde{\Pi}_{\mathsmaller{\mathrm{2D/g}}}(x)\to1$.

Below, we study the asymptotic behavior of Eqs.~(\ref{transresistivity_final}) and~(\ref{transresistivity_integral}) in three different limits corresponding to small and large values of $q_{\mathsmaller{\mathrm{TF}}}^{\mathsmaller{\mathrm{2D/g}}}d$. Here, we note that Eqs.~(\ref{transresistivity_final}) and~(\ref{transresistivity_integral}) have been derived under the assumption that the nonlinear susceptibilities can be approximated by Eqs.~(\ref{nonlinear_susceptibility_2DEG}) and~(\ref{nonlinear_susceptibility_graphene}). As shown in Ref.~\onlinecite{Narozhny2012:PRB}, however, this is not the case for the limit $d=0$, where the Fermi energy is no longer the largest scale of the system and Eq.~(\ref{nonlinear_susceptibility_graphene}) is not a good approximation for the nonlinear susceptibility in graphene [the same is also true for the nonlinear susceptibility of the 2DEG given by Eq.~(\ref{nonlinear_susceptibility_2DEG})]. Only for weak interaction strength, Eq.~(\ref{nonlinear_susceptibility_graphene}) can describe the nonlinear susceptibility for small, but finite $d$ reasonably well [For $d=0$, the transformation in the integral of Eq.~(\ref{transresistivity_integral}) cannot be used.].\cite{Carrega2012:arXiv,Amorim2012:arXiv} Thus, the limit of small interlayer separation presented in the following should be understood in this way.

\subsubsection{Small interlayer separation limit $(q_{\mathsmaller{\mathrm{TF}}}^{\mathsmaller{\mathrm{2D/g}}}d,k_{\mathsmaller{\mathrm{F}}}^{\mathsmaller{\mathrm{2D/g}}}d\ll1)$}

In this case, the integration in Eq.~(\ref{transresistivity_integral}) is restricted by the upper boundary $x_0=\mathrm{min}\left(2k_{\mathsmaller{\mathrm{F}}}^{\mathsmaller{\mathrm{2D}}}d,2k_{\mathsmaller{\mathrm{F}}}^{\mathsmaller{\mathrm{g}}}d\right)$ and we obtain
\begin{widetext}
\begin{equation}\label{transresistivity_asymptotic_ssd}
\rho_\mathsmaller{\mathrm{D}}=-\frac{h}{e^2}\frac{\left(k_\mathsmaller{\mathrm{B}}T\right)^2}{\epsilon_{\mathsmaller{\mathrm{F}}}^{\mathsmaller{\mathrm{2D}}}\epsilon_{\mathsmaller{\mathrm{F}}}^{\mathsmaller{\mathrm{g}}}}\frac{q_{\mathsmaller{\mathrm{TF}}}^{\mathsmaller{\mathrm{2D}}}q_{\mathsmaller{\mathrm{TF}}}^{\mathsmaller{\mathrm{g}}}}{k_{\mathsmaller{\mathrm{F}}}^{\mathsmaller{\mathrm{2D}}}k_{\mathsmaller{\mathrm{F}}}^{\mathsmaller{\mathrm{g}}}}\frac{\pi}{12}\Bigl[f\left(k_{\mathsmaller{\mathrm{F}}}^{\mathsmaller{\mathrm{2D}}},k_{\mathsmaller{\mathrm{F}}}^{\mathsmaller{\mathrm{g}}},q_{\mathsmaller{\mathrm{TF}}}^{\mathsmaller{\mathrm{2D}}}+q_{\mathsmaller{\mathrm{TF}}}^{\mathsmaller{\mathrm{g}}}\right)-g\left(k_{\mathsmaller{\mathrm{F}}}^{\mathsmaller{\mathrm{2D}}},k_{\mathsmaller{\mathrm{F}}}^{\mathsmaller{\mathrm{g}}},q_{\mathsmaller{\mathrm{TF}}}^{\mathsmaller{\mathrm{2D}}},q_{\mathsmaller{\mathrm{TF}}}^{\mathsmaller{\mathrm{g}}}\right)\sqrt{k_{\mathsmaller{\mathrm{F}}}^{\mathsmaller{\mathrm{2D}}}k_{\mathsmaller{\mathrm{F}}}^{\mathsmaller{\mathrm{g}}}}w+\mathcal{O}\left(k_{\mathsmaller{\mathrm{F}}}^{\mathsmaller{\mathrm{2D}}}k_{\mathsmaller{\mathrm{F}}}^{\mathsmaller{\mathrm{g}}}w^2\right)\Bigr],
\end{equation}
where
\begin{equation}\label{transresistivity_asymptotic_ssd_int}
f\left(k_{\mathsmaller{\mathrm{F}}}^{\mathsmaller{\mathrm{2D}}},k_{\mathsmaller{\mathrm{F}}}^{\mathsmaller{\mathrm{g}}},q_{\mathsmaller{\mathrm{TF}}}\right)=\int\limits_{0}^{y_0}\frac{y}{\left[y+q_{\mathsmaller{\mathrm{TF}}}/(2\sqrt{k_{\mathsmaller{\mathrm{F}}}^{\mathsmaller{\mathrm{2D}}}k_{\mathsmaller{\mathrm{F}}}^{\mathsmaller{\mathrm{g}}}})\right]^2}\sqrt{\frac{1-(k_{\mathsmaller{\mathrm{F}}}^{\mathsmaller{\mathrm{2D}}}/k_{\mathsmaller{\mathrm{F}}}^{\mathsmaller{\mathrm{g}}})y^2}{1-(k_{\mathsmaller{\mathrm{F}}}^{\mathsmaller{\mathrm{g}}}/k_{\mathsmaller{\mathrm{F}}}^{\mathsmaller{\mathrm{2D}}})y^2}}
\end{equation}
and
\begin{equation}\label{transresistivity_asymptotic_ssd_int_QWcor}
\begin{aligned}
g\left(k_{\mathsmaller{\mathrm{F}}}^{\mathsmaller{\mathrm{2D}}},k_{\mathsmaller{\mathrm{F}}}^{\mathsmaller{\mathrm{g}}},q_{\mathsmaller{\mathrm{TF}}}^{\mathsmaller{\mathrm{2D}}},q_{\mathsmaller{\mathrm{TF}}}^{\mathsmaller{\mathrm{g}}}\right)=\int\limits_{0}^{y_0}\frac{y\left\{24\pi^2y^2+2\left[6\pi^2r_{\mathsmaller{\mathrm{g}}}+\left(2\pi^2+15\right)r_{\mathsmaller{\mathrm{2D}}}\right]y+\left(8\pi^2+15\right)r_{\mathsmaller{\mathrm{2D}}}r_{\mathsmaller{\mathrm{g}}}\right\}}{12\pi^2\left[y+\left(r_{\mathsmaller{\mathrm{2D}}}+r_{\mathsmaller{\mathrm{g}}}\right)/2\right]^3}\sqrt{\frac{1-(k_{\mathsmaller{\mathrm{F}}}^{\mathsmaller{\mathrm{2D}}}/k_{\mathsmaller{\mathrm{F}}}^{\mathsmaller{\mathrm{g}}})y^2}{1-(k_{\mathsmaller{\mathrm{F}}}^{\mathsmaller{\mathrm{g}}}/k_{\mathsmaller{\mathrm{F}}}^{\mathsmaller{\mathrm{2D}}})y^2}}
\end{aligned}
\end{equation}
\end{widetext}
with $y_0=\mathrm{min}\left(\sqrt{k_{\mathsmaller{\mathrm{F}}}^{\mathsmaller{\mathrm{2D}}}/k_{\mathsmaller{\mathrm{F}}}^{\mathsmaller{\mathrm{g}}}},\sqrt{k_{\mathsmaller{\mathrm{F}}}^{\mathsmaller{\mathrm{g}}}/k_{\mathsmaller{\mathrm{F}}}^{\mathsmaller{\mathrm{2D}}}}\right)$ and $r_{\mathsmaller{\mathrm{2D/g}}}=q_{\mathsmaller{\mathrm{TF}}}^{\mathsmaller{\mathrm{2D/g}}}/\sqrt{k_{\mathsmaller{\mathrm{F}}}^{\mathsmaller{\mathrm{2D}}}k_{\mathsmaller{\mathrm{F}}}^{\mathsmaller{\mathrm{g}}}}$. Equation~(\ref{transresistivity_asymptotic_ssd}) shows that in the small interlayer separation limit the transresistivity does not depend on $d$. Such a behavior has also been found in graphene-graphene double-layer structures.\cite{Carrega2012:arXiv,Amorim2012:arXiv} Moreover, $g\left(k_{\mathsmaller{\mathrm{F}}}^{\mathsmaller{\mathrm{2D}}},k_{\mathsmaller{\mathrm{F}}}^{\mathsmaller{\mathrm{g}}},q_{\mathsmaller{\mathrm{TF}}}^{\mathsmaller{\mathrm{2D}}},q_{\mathsmaller{\mathrm{TF}}}^{\mathsmaller{\mathrm{g}}}\right)>0$ and thus the transresistivity is reduced for finite widths of the quantum well. Equations~(\ref{transresistivity_asymptotic_ssd_int}) and~(\ref{transresistivity_asymptotic_ssd_int_QWcor}) have --- in general --- to be computed numerically. However, for certain limiting cases analytical formulas can be derived for which we refer to Appendix~\ref{Sec:AppendixLimitingCases}. In particular, when the particle densities are such that $n=n_{\mathsmaller{\mathrm{g}}}=2n_{\mathsmaller{\mathrm{2D}}}$ we obtain, in the leading order of $1/n$, $\rho_\mathsmaller{\mathrm{D}} \propto 1/n^{2}$. This dependence can be seen as an intermediate behavior when compared to the results expected for 2DEG-2DEG ($\rho_\mathsmaller{\mathrm{D}} \propto 1/n^{3}$) and graphene-graphene ($\rho_\mathsmaller{\mathrm{D}} \propto 1/n$) double-layer structures when the particle density is equal in both layers.\cite{Narozhny2012:PRB,Carrega2012:arXiv,Amorim2012:arXiv}

\subsubsection{Intermediate limit $(q_{\mathsmaller{\mathrm{TF}}}^{\mathsmaller{\mathrm{2D/g}}}d\ll1\ll k_{\mathsmaller{\mathrm{F}}}^{\mathsmaller{\mathrm{2D/g}}}d)$}

Whereas the limit considered above requires small interlayer distances, $d$ must not be too small for the limit $q_{\mathsmaller{\mathrm{TF}}}^{\mathsmaller{\mathrm{2D/g}}}d\ll1,k_{\mathsmaller{\mathrm{F}}}^{\mathsmaller{\mathrm{2D/g}}}d\gg1$. It is difficult, however, to reach this limit experimentally because in graphene $q_{\mathsmaller{\mathrm{TF}}}^{\mathsmaller{\mathrm{g}}}\propto k_{\mathsmaller{\mathrm{F}}}^{\mathsmaller{\mathrm{g}}}/\overline{\kappa}$. Thus, this limit can only be reached if $\overline{\kappa}$ is very large.

Since $k_{\mathsmaller{\mathrm{F}}}^{\mathsmaller{\mathrm{2D/g}}}d\gg1$, the integral in Eq.~(\ref{transresistivity_integral}) is practically restricted by the Coulomb interaction with the main contribution arising from values $x=qd\lesssim1$. Therefore, we can approximate $\Lambda_{\mathsmaller{\mathrm{2D/g}}}(x/d)$ by $\Lambda_{\mathsmaller{\mathrm{2D/g}}}(x/d)\to1$ and find
\begin{equation}\label{transresistivity_asymptotic_sd}
\begin{aligned}
\rho_\mathsmaller{\mathrm{D}}=&-\frac{h}{e^2}\frac{\left(k_\mathsmaller{\mathrm{B}}T\right)^2}{\epsilon_{\mathsmaller{\mathrm{F}}}^{\mathsmaller{\mathrm{2D}}}\epsilon_{\mathsmaller{\mathrm{F}}}^{\mathsmaller{\mathrm{g}}}}\frac{q_{\mathsmaller{\mathrm{TF}}}^{\mathsmaller{\mathrm{2D}}}q_{\mathsmaller{\mathrm{TF}}}^{\mathsmaller{\mathrm{g}}}}{k_{\mathsmaller{\mathrm{F}}}^{\mathsmaller{\mathrm{2D}}}k_{\mathsmaller{\mathrm{F}}}^{\mathsmaller{\mathrm{g}}}}\frac{\pi}{12}\\
&\times\left\{\ln\left[\frac{1+\left(q_{\mathsmaller{\mathrm{TF}}}^{\mathsmaller{\mathrm{2D}}}+q_{\mathsmaller{\mathrm{TF}}}^{\mathsmaller{\mathrm{g}}}\right)d}{\left(q_{\mathsmaller{\mathrm{TF}}}^{\mathsmaller{\mathrm{2D}}}+q_{\mathsmaller{\mathrm{TF}}}^{\mathsmaller{\mathrm{g}}}\right)d}\right]-\frac{w}{d}+\mathcal{O}\left[(w/d)^2\right]\right\},
\end{aligned}
\end{equation}
which has been expanded in powers of $w/d$. From Eq.~(\ref{transresistivity_asymptotic_sd}), we obtain $\rho_\mathsmaller{\mathrm{D}}\sim\ln[(\alpha\sqrt{\pi n_\mathsmaller{\mathrm{g}}}+q_{\mathsmaller{\mathrm{TF}}}^{\mathsmaller{\mathrm{2D}}})d]/\sqrt{n^{3}_{\mathsmaller{\mathrm{2D}}}n_{\mathsmaller{\mathrm{g}}}}$ for the dependence of $\rho_\mathsmaller{\mathrm{D}}$ on the carrier densities, with $\alpha=4e^{2}/(\overline{\kappa}\hbar v_{\mathsmaller{\mathrm{F}}}^{\mathsmaller{\mathrm{g}}})$.

\subsubsection{Large interlayer separation limit $(q_{\mathsmaller{\mathrm{TF}}}^{\mathsmaller{\mathrm{2D/g}}}d,k_{\mathsmaller{\mathrm{F}}}^{\mathsmaller{\mathrm{2D/g}}}d\gg1)$}

As for the limit of intermediate interlayer separation above, the main contribution to Eq.~(\ref{transresistivity_integral}) arises for $x\lesssim1$ and we can approximate $\Lambda_{\mathsmaller{\mathrm{2D/g}}}(x/d)\to1$. In this case, the values contributing to Eq.~(\ref{transresistivity_integral}) satisfy $x\ll y_{\mathsmaller{\mathrm{2D/g}}}/\overline{\kappa}$ and we obtain
\begin{equation}\label{transresistivity_asymptotic_ld}
\begin{aligned}
\rho_\mathsmaller{\mathrm{D}}=&-\frac{h}{e^2}\frac{\left(k_{\mathsmaller{B}}T\right)^2}
{\epsilon_{\mathsmaller{\mathrm{F}}}^{\mathsmaller{\mathrm{2D}}}\epsilon_{\mathsmaller{\mathrm{F}}}^{\mathsmaller{\mathrm{g}}}
\left(k_{\mathsmaller{\mathrm{F}}}^{\mathsmaller{\mathrm{2D}}}d\right)
\left(k_{\mathsmaller{\mathrm{F}}}^{\mathsmaller{\mathrm{g}}}d\right)
\left(q_{\mathsmaller{\mathrm{TF}}}^{\mathsmaller{\mathrm{2D}}}d\right)
\left(q_{\mathsmaller{\mathrm{TF}}}^{\mathsmaller{\mathrm{g}}}d\right)}
\frac{\pi\zeta(3)}{32}\\
&\times\Biggl\{1-\frac{\left(720\pi^2+1350\right)\zeta(3)-\pi^4\left(4\pi^2-15\right)}{540\pi^2\zeta(3)}\frac{w}{d}\\
&\quad\quad+\mathcal{O}\left[(w/d)^2\right]\Biggr\}.
\end{aligned}
\end{equation}
Consequently, the dependence of $\rho_\mathsmaller{\mathrm{D}}$ on the carrier densities is given by $\rho_\mathsmaller{\mathrm{D}}\sim 1/[(n_{\mathsmaller{\mathrm{2D}}}n_{\mathsmaller{\mathrm{g}}})^{3/2}d^{4}]$, which is the same asymptotic behavior as one would expect in a double-layer structure consisting of two 2DEGs or one consisting of two graphene layers. Only when higher order terms in the series expansion of $\Lambda_{\mathsmaller{\mathrm{2D/g}}}(x/d)$ [see Eqs.~(\ref{g_2DEG}) and (\ref{g_graphene})] are taken into account, one can find a difference in the asymptotic behavior (see Sec.~\ref{Sec:Corrections}).

The asymptotic behavior of the transresistivity as a function of the carrier densities is summarized in Table \ref{tab:comparison}, where, for comparison, the results corresponding to massive-massive and massless-massless systems have also been included.

\subsection{Numerical calculations}
We have performed numerical calculations using Eqs.~(\ref{transresistivity_final}) and~(\ref{transresistivity_integral}) for two different structures, air/graphene/Al$_2$O$_3$/GaAs/AlGaAs and air/graphene/SiO$_2$/GaAs/AlGaAs, in which graphene plays the role of the active layer and the 2DEG formed in the GaAs quantum well constitutes the passive one. The two structures differ in the materials conforming the spacer, Al$_2$O$_3$ in the former and SiO$_2$ in the later case, which possess different dielectric constants, $\kappa_2=9.1$ and $\kappa_2=3.9$, respectively. The remaining system parameters used in the evaluation of the transresistivity are $\kappa_1=12.9$, $\kappa_3=1$, $\kappa_{\mathsmaller{\mathrm{2D}}}=12.9$, and the electron effective mass in GaAs, $m_{\mathsmaller{\rm GaAs}}^{\ast}=0.063\;m_0$. Here, $m_0$ represents the bare electron mass.

\begin{figure}[t]
\centering
\includegraphics*[width=0.95\columnwidth,angle=0]{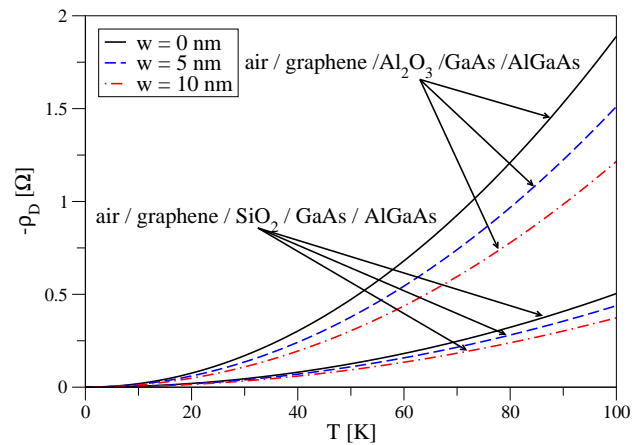}
\caption{(color online). Dependence of the transresistivity on the temperature, $T$, for air/graphene/Al$_2$O$_3$/GaAs/AlGaAs and air/graphene/SiO$_2$/GaAs/AlGaAs structures with an interlayer distance of $d=20$ nm for different widths of the GaAs quantum well ($w=0,5,10$ nm). The electronic densities of graphene and GaAs have been set to the values $n_{\rm g}=1.0\times10^{12}$ cm$^{-2}$ and $n_{\rm GaAs}=1.0\times10^{11}$ cm$^{-2}$, respectively.}
\label{fig:temperature}
\end{figure}

The temperature dependence of the transresistivity in the two considered structures is shown in Fig.~\ref{fig:temperature} for different widths of the GaAs quantum well. The interlayer distance is $d=20$ nm and the densities $n_{\rm GaAs}=1.0\times10^{11}$ cm$^{-2}$ and $n_{\rm g}=1.0\times10^{12}$ cm$^{-2}$. Here, the densities chosen for each layer reflect the possibility that for double-layer systems consisting of 2DEGs and graphene one might be able to have a combination of rather different densities in both layers, a case difficult to achieve if both layers consist of the same material. As can be seen from Eq.~(\ref{transresistivity_final}) and Fig.~\ref{fig:temperature}, $-\rho_\mathsmaller{\mathrm{D}}\sim T^2$. This is the same temperature-dependence found for the transresistivity in ballistic 2DEG-2DEG bilayers.\cite{Gramila1991:PRL,Zheng1993:PRB,Jauho1993:PRB,Kamenev1995:PRB,Flensberg1995:PRB} One can also appreciate in Fig.~\ref{fig:temperature} that, for a given temperature, the smaller the well width, the larger the size of the resistivity. This is a general behavior, which, according to Eqs.~(\ref{transresistivity_asymptotic_ssd}), (\ref{transresistivity_asymptotic_sd}), and (\ref{transresistivity_asymptotic_ld}), occurs in both the $q_{\mathsmaller{\mathrm{TF}}}^{\mathsmaller{\mathrm{2D/g}}}d\ll 1$ and the $q_{\mathsmaller{\mathrm{TF}}}^{\mathsmaller{\mathrm{2D/g}}}d\gg 1$ limits. Note, however, that for the set of parameters considered in Fig.~\ref{fig:temperature}, which corresponds more to the limit $q_{\mathsmaller{\mathrm{TF}}}^{\mathsmaller{\mathrm{2D/g}}}d\gg 1$, both $-\rho_\mathsmaller{\mathrm{D}}$ and its changes with the well width are larger in the air/graphene/Al$_2$O$_3$/GaAs/AlGaAs system than in air/graphene/SiO$_2$/GaAs/AlGaAs. This can be qualitatively understood by noting that in such a limit both $-\rho_{D}$ and $-\partial \rho_{D}/\partial w$ are proportional to $\overline{\kappa}^2$ [see Eq.~(\ref{transresistivity_asymptotic_ld}), where $\overline{\kappa}$ enters via the screened Thomas-Fermi wavevectors]. Consequently, since $\kappa_2$ (and therefore $\overline{\kappa}$) is larger in air/graphene/Al$_2$O$_3$/GaAs/AlGaAs than in air/graphene/SiO$_2$/GaAs/AlGaAs, both the absolute size of the transresistivity and its changes with $w$ are expected to be larger in the former structure compared to the later one, as is indeed seen in Fig.~\ref{fig:temperature}.

\begin{figure}[t]
\centering
\includegraphics*[width=0.95\columnwidth,angle=0]{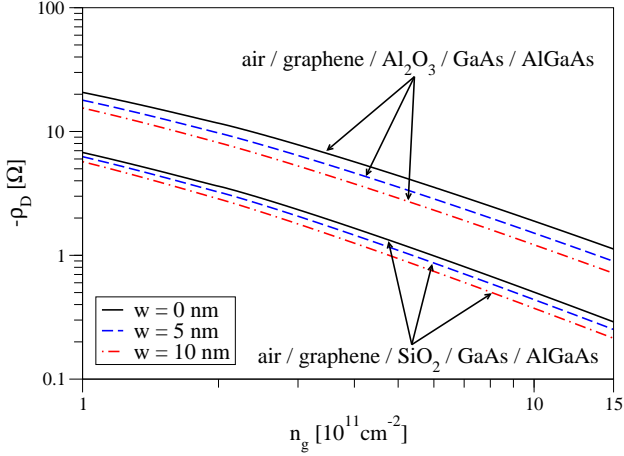}
\caption{(color online). Dependence of the transresistivity on the electronic density in graphene, $n_\mathsmaller{\mathrm{g}}$, at $T=100$ K for air/graphene/Al$_2$O$_3$/GaAs/AlGaAs and air/graphene/SiO$_2$/GaAs/AlGaAs structures with an interlayer distance of $d=20$ nm and for different widths of the GaAs quantum well ($w=0,5,10$ nm). The electronic density of GaAs has been set to the value $n_{\rm GaAs}=1.0\times10^{11}$ cm$^{-2}$.}
\label{fig:density}
\end{figure}

The dependence of the transresistivity on the density in the graphene layer at $T=100$ K is shown in Fig.~\ref{fig:density} for graphene/Al$_2$O$_3$/GaAs/AlGaAs and graphene/SiO$_2$/GaAs/AlGaAs structures. Here, the interlayer distance is set at $d=20$ nm and the density in the GaAs layer at $n_{\rm GaAs}=1.0\times10^{11}$ cm$^{-2}$. We can fit the curves in Fig.~\ref{fig:density} to $-\rho_{D}\propto n_{\rm g}^\beta$ and extract values between $\beta\approx-1.19$ and $\beta\approx-1.30$, which are closer to $-3/2$ than to $-1/2$. This is consistent with the fact that, for the parameters chosen, we are approaching the limit of $q_{\mathsmaller{\mathrm{TF}}}^{\mathsmaller{\mathrm{2D/g}}}d\gg 1$.

Regarding the dependence of $-\rho_\mathsmaller{\mathrm{D}}$ on the well width, both Fig.~\ref{fig:temperature} and Fig.~\ref{fig:density} exhibit the same qualitative behavior: the smaller the well width, the larger the absolute value of the transresistivity.

\section{Drag resistivity in bilayer graphene/monolayer graphene systems}\label{Sec:bilayer_monolayer_graphene}

Apart from the 2DEG-graphene system considered in the previous section, the Coulomb drag between massless and massive fermions may also be realized in a double-layer structure consisting of bilayer and monolayer graphene isolated from each other by a spacer. Compared to the system investigated in Sec.~\ref{Sec:2DEG_graphene}, the quantum well of width $w$ containing the 2DEG is replaced by a sheet ($w=0$) of bilayer graphene. As before, monolayer graphene is assumed to be the active layer, while bilayer graphene is taken as the passive one. Likewise, the ballistic case is assumed for both layers and only low temperatures are considered. We restrict our analysis to the case in which both layers are electron doped. In such a case, bilayer graphene consists of an electron band with a parabolic dispersion as in Eq.~(\ref{dispersion-2deg}), but with the BLG effective mass $m_{\rm bg}^{\ast}=0.033\;m_{0}$. With the relaxation time in bilayer graphene at the Fermi level given by $\tau_\mathsmaller{\mathrm{bg}}$, for $T\to0$ one obtains
\begin{equation}\label{nonlinear_susceptibility_blg}
\Gamma_{\mathsmaller{\mathrm{bg}}}^{i}\left(\mathbf{q},\omega\right)=-\frac{4e\omega\tau_\mathsmaller{\mathrm{bg}}}{\hbar\pi v_{\mathsmaller{\mathrm{F}}}^{\mathsmaller{\mathrm{bg}}}}\frac{q_i}{q}\Lambda_{\mathsmaller{\mathrm{bg}}}(q)
\end{equation}
for the nonlinear susceptibility in the limit of low energies and long wavelengths, where
\begin{equation}\label{g_blg}
\begin{aligned}
\Lambda_{\mathsmaller{\mathrm{bg}}}(q)=&\left[\frac{1}{\sqrt{1-\left(\frac{q}{2k_{\mathsmaller{\mathrm{F}}}^\mathsmaller{\mathrm{bg}}}\right)^2}}-
\left(\frac{q}{k_{\mathsmaller{\mathrm{F}}}^\mathsmaller{\mathrm{bg}}}\right)^2\sqrt{1-\left(\frac{q}{2k_{\mathsmaller{\mathrm{F}}}^\mathsmaller{\mathrm{bg}}}\right)^2}\right]\\
&\times\Theta\left(2k_{\mathsmaller{\mathrm{F}}}^\mathsmaller{\mathrm{bg}}-q\right)
\end{aligned}
\end{equation}
and $k_{\mathsmaller{\mathrm{F}}}^\mathsmaller{\mathrm{bg}}$ and $v_{\mathsmaller{\mathrm{F}}}^\mathsmaller{\mathrm{bg}}$ denote the Fermi wavevector and velocity in bilayer graphene.

We can replace the quantities describing the 2DEG layer by the respective quantities of bilayer graphene and use the results from Sec.~\ref{Sec:2DEG_graphene}. This means that, aside from setting $w=0$, replacing the Fermi energy, the Fermi velocity, the Fermi and Thomas-Fermi wavevectors, as well as $\Lambda_{\mathsmaller{\mathrm{2D}}}$, the polarization function of the 2DEG in Eq.~(\ref{transresistivity_integral}) has to be replaced by the polarization function of bilayer graphene calculated in Ref.~\onlinecite{Sensarma2010:PRB}. However, in contrast to a 2DEG, one has to take into account the valley degeneracy of bilayer graphene, the net effect of which is an additional factor of 1/2 in Eqs.~(\ref{transresistivity_final}), (\ref{transresistivity_asymptotic_sd}), and (\ref{transresistivity_asymptotic_ld}).

Whereas the limiting cases of intermediate and large interlayer separation from Sec.~\ref{Sec:Asymptotics} do not depend on the exact form of $\Lambda_{\mathsmaller{2D/bg}}$ because of $k_{\mathsmaller{\mathrm{F}}}^\mathsmaller{2D/bg}d\gg1$ and can therefore be described by Eqs.~(\ref{transresistivity_asymptotic_sd}) and (\ref{transresistivity_asymptotic_ld}), respectively (and taking into account the additional factor of 1/2 for bilayer graphene-graphene systems as well as setting $w=0$), this cannot be done in the small interlayer separation limit. For this limit, we obtain
\begin{equation}\label{transresistivity_asymptotic_ssd_bg}
\begin{aligned}
\rho_\mathsmaller{\mathrm{D}}=&-\frac{h}{e^2}\frac{\left(k_\mathsmaller{\mathrm{B}}T\right)^2}{\epsilon_{\mathsmaller{\mathrm{F}}}^{\mathsmaller{\mathrm{bg}}}\epsilon_{\mathsmaller{\mathrm{F}}}^{\mathsmaller{\mathrm{g}}}}\frac{q_{\mathsmaller{\mathrm{TF}}}^{\mathsmaller{\mathrm{bg}}}q_{\mathsmaller{\mathrm{TF}}}^{\mathsmaller{\mathrm{g}}}}{k_{\mathsmaller{\mathrm{F}}}^{\mathsmaller{\mathrm{bg}}}k_{\mathsmaller{\mathrm{F}}}^{\mathsmaller{\mathrm{g}}}}\frac{\pi}{24}\Bigl[f\left(k_{\mathsmaller{\mathrm{F}}}^{\mathsmaller{\mathrm{bg}}},k_{\mathsmaller{\mathrm{F}}}^{\mathsmaller{\mathrm{g}}},q_{\mathsmaller{\mathrm{TF}}}^{\mathsmaller{\mathrm{bg}}}+q_{\mathsmaller{\mathrm{TF}}}^{\mathsmaller{\mathrm{g}}}\right)\\
&-f_1\left(k_{\mathsmaller{\mathrm{F}}}^{\mathsmaller{\mathrm{bg}}},k_{\mathsmaller{\mathrm{F}}}^{\mathsmaller{\mathrm{g}}},q_{\mathsmaller{\mathrm{TF}}}^{\mathsmaller{\mathrm{bg}}}+q_{\mathsmaller{\mathrm{TF}}}^{\mathsmaller{\mathrm{g}}}\right)\Bigr],
\end{aligned}
\end{equation}
where $f(k_{\mathsmaller{\mathrm{F}}}^{\mathsmaller{\mathrm{bg}}},k_{\mathsmaller{\mathrm{F}}}^{\mathsmaller{\mathrm{g}}},q_{\mathsmaller{\mathrm{TF}}})$ is given by Eq.~(\ref{transresistivity_asymptotic_ssd_int}) and
\begin{equation}\label{transresistivity_asymptotic_ssd_int2_bg}
\begin{aligned}
f_1&\left(k_{\mathsmaller{\mathrm{F}}}^{\mathsmaller{\mathrm{bg}}},k_{\mathsmaller{\mathrm{F}}}^{\mathsmaller{\mathrm{g}}},q_{\mathsmaller{\mathrm{TF}}}\right)=\\
&\frac{4k_{\mathsmaller{\mathrm{F}}}^{\mathsmaller{\mathrm{g}}}}{k_{\mathsmaller{\mathrm{F}}}^{\mathsmaller{\mathrm{bg}}}}\int\limits_{0}^{y_0}\frac{y^3\sqrt{\left[1-(k_{\mathsmaller{\mathrm{F}}}^{\mathsmaller{\mathrm{bg}}}/k_{\mathsmaller{\mathrm{F}}}^{\mathsmaller{\mathrm{g}}})y^2\right]\left[1-(k_{\mathsmaller{\mathrm{F}}}^{\mathsmaller{\mathrm{g}}}/k_{\mathsmaller{\mathrm{F}}}^{\mathsmaller{\mathrm{bg}}})y^2\right]}}{\left[y+q_{\mathsmaller{\mathrm{TF}}}/(2\sqrt{k_{\mathsmaller{\mathrm{F}}}^{\mathsmaller{\mathrm{bg}}}k_{\mathsmaller{\mathrm{F}}}^{\mathsmaller{\mathrm{g}}}})\right]^2}
\end{aligned}
\end{equation}
with $y_0=\mathrm{min}\left(\sqrt{k_{\mathsmaller{\mathrm{F}}}^{\mathsmaller{\mathrm{bg}}}/k_{\mathsmaller{\mathrm{F}}}^{\mathsmaller{\mathrm{g}}}},\sqrt{k_{\mathsmaller{\mathrm{F}}}^{\mathsmaller{\mathrm{g}}}/k_{\mathsmaller{\mathrm{F}}}^{\mathsmaller{\mathrm{bg}}}}\right)$. In general, Eq.~(\ref{transresistivity_asymptotic_ssd_int2_bg}) has to be computed numerically, but it is possible to derive analytical formulas for certain limiting cases (see Appendix~\ref{Sec:AppendixLimitingCases}).

Thus, the transresistivity due to the Coulomb drag between massive and massless fermions in 2DEG-graphene and BLG-MLG structures is characterized by the same generic expressions for the intermediate and large interlayer separation limits. For small interlayer distances on the other hand, there is an additional contribution in the case of the BLG-graphene system compared to the 2DEG-graphene system [see the massless-massive system in Tab.~\ref{tab:comparison}].

\begin{figure}[t]
\centering
\includegraphics*[width=0.95\columnwidth,angle=0]{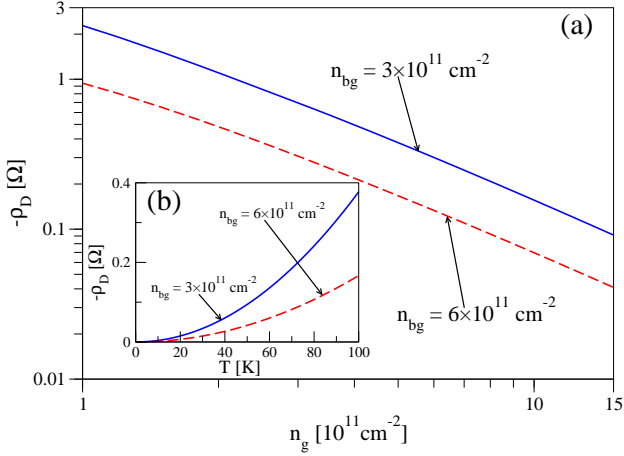}
\caption{(color online). (a) Dependence of the transresistivity of an air/graphene/Al$_2$O$_3$/bilayer graphene/SiO$_2$ structure on the electronic density of graphene ($n_\mathsmaller{\mathrm{g}}$), for different electronic densities in bilayer graphene ($n_\mathsmaller{\mathrm{bg}}$) at $T=100$ K. (b) Temperature dependence of the transresistivity for the same structure as in (a) at fixed graphene electronic density, $n_{\mathsmaller{\mathrm{g}}}=5\times10^{11}$ cm$^{-2}$. In both cases, (a) and (b), the interlayer distance is $d=20$ nm.}
\label{fig:BLG}
\end{figure}

Figure~\ref{fig:BLG} shows the dependence of the transresistivity in an air/graphene/Al$_2$O$_3$/bilayer graphene/SiO$_2$ structure on the electronic density of graphene ($n_\mathsmaller{\mathrm{g}}$) at $T=100$ K [Fig.~\ref{fig:BLG}~(a)] as well as on the temperature [Fig.~\ref{fig:BLG}~(b)] for $n_{\mathsmaller{\mathrm{g}}}=5\times10^{11}$ cm$^{-2}$ and different densities in bilayer graphene ($n_\mathsmaller{\mathrm{bg}}$). As anticipated above, the qualitative trends displayed in Fig.~\ref{fig:BLG} are similar to those shown in Figs.~\ref{fig:temperature} and~\ref{fig:density}. However, for the set parameters taken in Fig.~\ref{fig:BLG}, the system starts to approach the large interlayer separation limit and the absolute value of the drag resistivity in the air/graphene/Al$_2$O$_3$/bilayer graphene/SiO$_2$ appears to be smaller than in the 2DEG-graphene system. This behavior can be understood from Eq.~(\ref{transresistivity_asymptotic_ld}) by taking into account that the screened Thomas-Fermi wavevector in BLG is larger than in GaAs.

\section{Corrections to the large interlayer separation limit}\label{Sec:Corrections}
As mentioned in Sec.~\ref{Sec:2DEG_graphene} and shown in Tab.~\ref{tab:comparison}, the asymptotic behavior of the transresistivity in a massless-massive double-layer system in the limit of $q_{\mathsmaller{\mathrm{TF}}}^{\mathsmaller{\mathrm{a/p}}}d\gg1$ is identical to the behavior in massive-massive and massless-massless systems ($a/p$ denote the active/pasive layers). Only when higher order terms in the series expansion of $\Lambda_{\mathsmaller{\mathrm{a/p}}}(x/d)$ [see Eqs.~(\ref{g_2DEG}), (\ref{g_graphene}), and (\ref{g_blg})] are taken into account, one can find a difference in the asymptotic behavior.

In general, we find that the leading correction (for $w=0$) to Eq.~(\ref{transresistivity_asymptotic_ld}) is given by
\begin{equation}\label{general_system_correction}
\begin{aligned}
\Delta\rho_\mathsmaller{\mathrm{D}}\approx&-\frac{h}{e^2}
\frac{5\pi\zeta(5)
\left(k_{\mathsmaller{B}}T\right)^2
\left[c_{\mathsmaller{\mathrm{p}}}(k_{\mathsmaller{\mathrm{F}}}^{\mathsmaller{\mathrm{a}}}d)^2+c_{\mathsmaller{\mathrm{a}}}(k_{\mathsmaller{\mathrm{F}}}^{\mathsmaller{\mathrm{p}}}d)^2\right]}
{256g_{\mathsmaller{\mathrm{a}}}g_{\mathsmaller{\mathrm{p}}}\epsilon_{\mathsmaller{\mathrm{F}}}^{\mathsmaller{\mathrm{a}}}\epsilon_{\mathsmaller{\mathrm{F}}}^{\mathsmaller{\mathrm{p}}}
\left(k_{\mathsmaller{\mathrm{F}}}^{\mathsmaller{\mathrm{a}}}d\right)^{3}
\left(k_{\mathsmaller{\mathrm{F}}}^{\mathsmaller{\mathrm{p}}}d\right)^{3}
\left(q_{\mathsmaller{\mathrm{TF}}}^{\mathsmaller{\mathrm{a}}}d\right)
\left(q_{\mathsmaller{\mathrm{TF}}}^{\mathsmaller{\mathrm{p}}}d\right)},
\end{aligned}
\end{equation}
where $a/p$, $k_{\mathsmaller{\mathrm{F}}}^\mathsmaller{\mathrm{a/p}}$, and $q_{\mathsmaller{\mathrm{F}}}^\mathsmaller{\mathrm{a/p}}$ denote the active/passive layers and their respective Fermi and screened Thomas-Fermi wavevectors. The parameters $c_{\mathsmaller{\mathrm{a/p}}}$ and $g_{\mathsmaller{\mathrm{a/p}}}$ are specific of the system comprising the active/passive layers, with $c_{\mathsmaller{\mathrm{2D}}}=1$, $c_{\mathsmaller{\mathrm{g}}}=-1$, and $c_{\mathsmaller{\mathrm{bg}}}=-7$, as well as $g_{\mathsmaller{\mathrm{2D}}}=g_{\mathsmaller{\mathrm{g}}}=1$ and $g_{\mathsmaller{\mathrm{bg}}}=2$.

From Eq.~(\ref{general_system_correction}), we find $\Delta\rho_\mathsmaller{\mathrm{D}}\sim(n_{\mathsmaller{\mathrm{g}}}-2n_{\mathsmaller{\mathrm{2D}}})/[(n_{\mathsmaller{\mathrm{2D}}}n_{\mathsmaller{\mathrm{g}}})^{5/2}d^{6}]$ for the density dependence in 2DEG-graphene systems in contrast to $\Delta\rho_\mathsmaller{\mathrm{D}}\sim \mp(n_{\mathsmaller{\mathrm{a}}}+n_{\mathsmaller{\mathrm{p}}})/[(n_{\mathsmaller{\mathrm{a}}}n_{\mathsmaller{\mathrm{p}}})^{5/2}d^{6}]$ in 2DEG-2DEG ($-$) and graphene-graphene ($+$) systems. In particular, under the condition of $n=n_{\mathsmaller{\mathrm{g}}}=2n_{\mathsmaller{\mathrm{2D}}}$ the correction to the drag vanishes for the 2DEG-graphene system but remains finite, with the asymptotic behavior $\Delta\rho_\mathsmaller{\mathrm{D}}\sim 1/(n^{4}d^6)$, for the 2DEG-2DEG and graphene-graphene systems. Similarly, Eq.~(\ref{general_system_correction}) can be used to describe deviations from the asymptotic behavior for 2DEG-BLG, BLG-BLG, and BLG-(monolayer) graphene systems.

In the limit $q_{\mathsmaller{\mathrm{TF}}}^{\mathsmaller{\mathrm{a/p}}}d\gg1$, the drag correction is small, in agreement with the trend described by Eq.~(\ref{general_system_correction}). For the set of parameters considered here (which correspond to a region close, but still not in such a limit), we have found from our numerical calculations that while the drag correction in air/graphene/Al$_2$O$_3$/GaAs/AlGaAs and air/graphene/SiO$_2$/GaAs/AlGaAs turns out to be still small (a few percent of the total drag), it becomes relevant for the case of air/graphene/Al$_2$O$_3$/bilayer graphene/SiO$_2$, in which it represents about $30\%$ of the total transresistivity.

For the case of air/graphene/Al$_2$O$_3$/bilayer graphene/SiO$_2$ with the parameters used in Fig.~\ref{fig:BLG}, the limit $q_{\mathsmaller{\mathrm{TF}}}^{\mathsmaller{\mathrm{a/p}}}d\gg1$ is reached when the interlayer distance is increased to values $d\gtrsim 80\textrm{ nm}$. In such a limit, we found a very good agreement between our numerical calculations and Eq.~(\ref{general_system_correction}). Our calculations indicate that the drag correction decreases from $12-13\%$ to $1\%$ (at $n_{\rm g}=1.5\times10^{12}$ cm$^{-2}$) of the total transresistivity when the interlayer distance is increased from $d=20\textrm{ nm}$ to $d= 80\textrm{ nm}$.

\section{Conclusions}\label{Sec:Conclusions}
In this manuscript, we have studied the Coulomb drag at low temperatures in a double-layer structure composed of a 2DEG and a graphene layer, both of which were treated as being in the Boltzmann regime. We have written down a formula to describe the transresistivity of such a system at low temperatures and have analyzed the temperature and density dependence of this formula both analytically and numerically. Analytical formulae have been derived to describe the asymptotic behavior in both, the small and large interlayer separation limits and compared to the respective behavior in massive-massive as well as massless-massless systems. It has been found that for $q_{\mathsmaller{\mathrm{TF}}}^{\mathsmaller{\mathrm{a/p}}}d\ll1$ each system, massive-massive, massless-massless, and massless-massive, possesses a different dependence on the  carrier densities, whereas the three systems share the same behavior in the dominant contribution to $\rho_\mathsmaller{\mathrm{D}}$ for $q_{\mathsmaller{\mathrm{TF}}}^{\mathsmaller{\mathrm{a/p}}}d\gg1$. Only looking at higher-order corrections allows us to distinguish between the different systems in this regime. Furthermore, the effect of a finite width of the quantum well in which the 2DEG is formed has been investigated and we have seen that with increasing well width the absolute value of the transresistivity is reduced.
Finally, we have also studied a BLG-graphene system, which we found to be qualitatively similar to a 2DEG-graphene system in the large interlayer separation limit, but different in the limit of small interlayer separation.

\acknowledgments{We are grateful to J. Fabian for useful hints and discussions. This work was supported by the DFG via SFB 689 and GRK 1570.}

\appendix
\section{Bare Coulomb potential}
\label{Sec:AppendixCoulomb}

The bare Coulomb potentials can be obtained from the Poisson equation, which in cylindrical coordinates $\left(\bm{\rho},z\right)$ reads as
\begin{equation}\label{Poisson_real}
\nabla\left[\kappa(z)\nabla\phi(\bm{\rho}-\bm{\rho}';z,z')\right]=4\pi e\delta\left(\bm{\rho}-\bm{\rho}'\right)\delta\left(z-z'\right)
\end{equation}
for a point charge located in a geometry as shown in Fig.~\ref{fig:Graphene_vs_2DEG}. Here, the relative dielectric constant is given by
\begin{equation}\label{kappa_def}
\kappa(z)=\left\{\begin{array}{ll}
            \kappa_3 & \mathrm{for}\quad z>d+w\\
            \kappa_2 & \mathrm{for}\quad w<z<d+w\\
            \kappa_{\mathsmaller{\mathrm{2D}}} & \mathrm{for}\quad 0<z<w\\
            \kappa_1 & \mathrm{for}\quad z<0.\\
            \end{array}\right.
\end{equation}
Introducing the Fourier transform of $\phi$ with respect to the in-plane coordinates $\bm{\rho}$ and insertion in Eq.~(\ref{Poisson_real}) yields
\begin{equation}\label{Poisson_momentum}
\frac{\d}{\d z}\left[\kappa(z)\frac{\d\phi(\mathbf{q};z,z')}{\d z}\right]-\kappa(z)q^2\phi(\mathbf{q};z,z')=4\pi e\delta\left(z-z'\right).
\end{equation}
This equation is solved for each region given in Eq.~(\ref{kappa_def}) (and each combination of $z$ and $z'$) separately and we require the global solution to be continuous and its derivative to be piecewise continuous with a jump of $4\pi e$ at $z=z'$.

Having determined the potential $\phi(\mathbf{q};z,z')$ in this way, the bare Coulomb potential can be calculated from
\begin{equation}\label{Coulomb_potential}
U_{\mathsmaller{ij}}^{(0)}\left(\mathbf{q}\right)=-e\int\limits_{-\infty}^{\infty}\d z\int\limits_{-\infty}^{\infty}\d z'\phi(\mathbf{q};z,z')\left|\chi_i(z)\right|^2\left|\chi_j(z')\right|^2,
\end{equation}
where $\chi_{i/j}(z)$ describes the localization in $z$ direction of a particle located in the 2DEG ($i=2D$) or graphene ($i=g$) layers. For graphene, we assume the electrons to be perfectly localized and therefore
\begin{equation}\label{wavefunction_graphene}
\left|\chi_{\mathsmaller{\mathrm{g}}}(z)\right|^2=\delta\left(z-d-w\right).
\end{equation}
The transversal wavefunction of an electron located in the 2DEG quantum well, on the other hand, is assumed to be given by that of the ground state of an infinite one-dimensional potential well,
\begin{equation}\label{wavefunction_2DEG}
\left|\chi_{\mathsmaller{\mathrm{2D}}}(z)\right|^2=\frac{2}{w}\sin^2\left(\frac{\pi z}{w}\right)\Theta\left(w-z\right)\Theta\left( z\right),
\end{equation}
that is, we assume that only the lowest quantum well subband is occupied.
From the solution of Eq.~(\ref{Poisson_momentum}), $\phi(\mathbf{q};z,z')$, and Eqs.~(\ref{Coulomb_potential})-(\ref{wavefunction_2DEG}) we find the interlayer potential $U_{\mathsmaller{\mathrm{2Dg}}}^{(0)}\left(\mathbf{q}\right)=(8\pi e^2/q)f_{\mathsmaller{\mathrm{2Dg}}}\left(qd,qw\right)$ ($i=2D$ and $j=g$ or vice versa) and the intralayer potentials $U_{\mathsmaller{\mathrm{2D/g}}}^{(0)}\left(\mathbf{q}\right)=(4\pi e^2/q)f_{\mathsmaller{\mathrm{2D/g}}}\left(qd,qw\right)$ ($i=j=\mathrm{2D}$ and $i=j=\mathrm{g}$) with the form factors
\begin{equation}\label{f_inter}
f_{\mathsmaller{\mathrm{2Dg}}}\left(x,y\right)=\frac{2\pi^2\kappa_2\left\{\kappa_1\left[\cosh(y)-1\right]+\kappa_{\mathsmaller{\mathrm{2D}}}\sinh(y)\right\}}{y\left(y^2+4\pi^2\right)\mathcal{N}(x,y)},
\end{equation}
\begin{equation}\label{f_intra_g}
\begin{aligned}
f_{\mathsmaller{\mathrm{g}}}\left(x,y\right)=&\frac{\kappa_2\cosh(x)\left[\kappa_1\sinh(y)+\kappa_{\mathsmaller{\mathrm{2D}}}\cosh(y)\right]}{\mathcal{N}(x,y)}\\
&+\frac{\kappa_{\mathsmaller{\mathrm{2D}}}\sinh(x)\left[\kappa_1\cosh(y)+\kappa_{\mathsmaller{\mathrm{2D}}}\sinh(y)\right]}{\mathcal{N}(x,y)},
\end{aligned}
\end{equation}
\begin{widetext}
\begin{equation}\label{f_intra_2D}
\begin{aligned}
f_{\mathsmaller{\mathrm{2D}}}\left(x,y\right)=&\frac{\kappa_1\kappa_2\left[\kappa_2\sinh(x)+\kappa_3\cosh(x)\right]\left\{64\pi^4\left[1-\cosh(y)\right]+y\left(y^2+4\pi^2\right)\left(3y^2+8\pi^2\right)\sinh(y)\right\}}{2\kappa_{\mathsmaller{\mathrm{2D}}}y^2\left(y^2+4\pi^2\right)^2\mathcal{N}(x,y)}\\
&+\frac{\left[\kappa_2\left(\kappa_1+\kappa_3\right)\cosh(x)+\left(\kappa_2^2+\kappa_1\kappa_3\right)\sinh(x)\right]\left[y\left(32\pi^4+20\pi^2y^2+3y^4\right)\cosh(y)-32\pi^4\sinh(y)\right]}{2y^2\left(y^2+4\pi^2\right)^2\mathcal{N}(x,y)}\\
&+\frac{\kappa_{\mathsmaller{\mathrm{2D}}}\left[\kappa_2\cosh(x)+\kappa_3\sinh(x)\right]y\left(y^2+4\pi^2\right)\left(3y^2+8\pi^2\right)\sinh(y)}{2y^2\left(y^2+4\pi^2\right)^2\mathcal{N}(x,y)},
\end{aligned}
\end{equation}
where
\begin{equation}\label{denominator}
\begin{aligned}
\mathcal{N}(x,y)=&\kappa_2\cosh(x)\left[\kappa_{\mathsmaller{\mathrm{2D}}}\left(\kappa_1+\kappa_3\right)\cosh(y)+\left(\kappa_1\kappa_3+\kappa^2_{\mathsmaller{\mathrm{2D}}}\right)\sinh(y)\right]\\
&+\sinh(x)\left[\kappa_{\mathsmaller{\mathrm{2D}}}\left(\kappa_2^2+\kappa_1\kappa_3\right)\cosh(y)+\left(\kappa_1\kappa_2^2+\kappa_3\kappa^2_{\mathsmaller{\mathrm{2D}}}\right)\sinh(y)\right].
\end{aligned}
\end{equation}
\end{widetext}
In the limit of $y\to0$ [$f_i(x)\equiv f_i(x,0)$], which describes the setup of Sec.~\ref{Sec:bilayer_monolayer_graphene}, we recover
\begin{equation}\label{f_inter_limit}
f_{\mathsmaller{\mathrm{2Dg}}}\left(x\right)=\frac{\kappa_2\e^x}{\left(\kappa_1-\kappa_2\right)\left(\kappa_2-\kappa_3\right)+\e^{2x}\left(\kappa_1+\kappa_2\right)\left(\kappa_2+\kappa_3\right)},
\end{equation}
\begin{equation}\label{f_intra_g_limit}
f_{\mathsmaller{\mathrm{g}}}\left(x\right)=\frac{\left(\kappa_2-\kappa_1\right)+\left(\kappa_1+\kappa_2\right)\e^{2x}}{\left(\kappa_1-\kappa_2\right)\left(\kappa_2-\kappa_3\right)+\e^{2x}\left(\kappa_1+\kappa_2\right)\left(\kappa_2+\kappa_3\right)},
\end{equation}
\begin{equation}\label{f_intra_2D_limit}
f_{\mathsmaller{\mathrm{2D}}}\left(x\right)=\frac{\left(\kappa_2-\kappa_3\right)+\left(\kappa_2+\kappa_3\right)\e^{2x}}{\left(\kappa_1-\kappa_2\right)\left(\kappa_2-\kappa_3\right)+\e^{2x}\left(\kappa_1+\kappa_2\right)\left(\kappa_2+\kappa_3\right)},
\end{equation}
consistent with the results found in Refs.~\onlinecite{Profumo2010:PRB,Badalyan2012:PRB}.

\section{Limiting cases for small interlayer distances}
\label{Sec:AppendixLimitingCases}
As noted in Secs.~\ref{Sec:2DEG_graphene} and~\ref{Sec:bilayer_monolayer_graphene}, the integrals defined by Eqs.~(\ref{transresistivity_asymptotic_ssd_int}),(\ref{transresistivity_asymptotic_ssd_int_QWcor}), and~(\ref{transresistivity_asymptotic_ssd_int2_bg}) have to be computed numerically, but it is possible to obtain analytical formulas for certain limiting cases. In the following, some of these cases are presented.

If $k_{\mathsmaller{\mathrm{F}}}^{\mathsmaller{m}}\approx k_{\mathsmaller{\mathrm{F}}}^{\mathsmaller{\mathrm{g}}}\equiv k_{\mathsmaller{\mathrm{F}}}$ ($m=2D/bg$), that is, for carrier densities $n_\mathsmaller{\mathrm{g}}\approx2n_\mathsmaller{\mathrm{2D}}$ in a 2DEG-graphene structure or $n_\mathsmaller{\mathrm{g}}\approx n_\mathsmaller{\mathrm{bg}}$ in a BLG-graphene structure, the integrands can be expanded around $k_{\mathsmaller{\mathrm{F}}}^{\mathsmaller{m}}/k_{\mathsmaller{\mathrm{F}}}^{\mathsmaller{\mathrm{g}}}=1$ and the integration of the lowest order contributions yields
\begin{equation}\label{f_equal}
f\left(k_{\mathsmaller{\mathrm{F}}},k_{\mathsmaller{\mathrm{F}}},q_{\mathsmaller{\mathrm{TF}}}\right)=\ln\left(\frac{1+\gamma}{\gamma}\right)-\frac{1}{1+\gamma}
\end{equation}
and
\begin{equation}\label{f1_equal}
\begin{aligned}
f_1\left(k_{\mathsmaller{\mathrm{F}}},k_{\mathsmaller{\mathrm{F}}},q_{\mathsmaller{\mathrm{TF}}}\right)=&4\gamma^2(3-5\gamma^2)\ln\left(\frac{1+\gamma}{\gamma}\right)\\
&+20\gamma^3-10\gamma^2-\frac{16}{3}\gamma+1,
\end{aligned}
\end{equation}
where $\gamma=q_{\mathsmaller{\mathrm{TF}}}/(2k_{\mathsmaller{\mathrm{F}}})$. For the massive-massless systems investigated in Secs.~\ref{Sec:2DEG_graphene} and~\ref{Sec:bilayer_monolayer_graphene}, the functions $f$ and $f_1$ have to be evaluated at $q_{\mathsmaller{\mathrm{TF}}}=q_{\mathsmaller{\mathrm{TF}}}^{\mathsmaller{\mathrm{g}}}+q_{\mathsmaller{\mathrm{TF}}}^{\mathsmaller{m}}$ [see Eqs.~(\ref{transresistivity_asymptotic_ssd}) and~(\ref{transresistivity_asymptotic_ssd_bg})] and thus $\gamma=\alpha/2+q_{\mathsmaller{\mathrm{TF}}}^{\mathsmaller{m}}/(2\sqrt{\pi n_\mathsmaller{\mathrm{g}}})$. Assuming that we are in the high-density limit (consistent with our usage of RPA), we can further simplify the transresistivity, given by Eqs.~(\ref{transresistivity_asymptotic_ssd}), (\ref{transresistivity_asymptotic_ssd_bg}), (\ref{f_equal}) and~(\ref{f1_equal}) if $k_{\mathsmaller{\mathrm{F}}}^{\mathsmaller{m}}=k_{\mathsmaller{\mathrm{F}}}^{\mathsmaller{\mathrm{g}}}$, by expanding it in powers of the density, from which we obtain $\rho_\mathsmaller{\mathrm{D}}\propto 1/n_{\mathsmaller{\mathrm{g}}}^2$. This is the formula used in Tab.~\ref{tab:comparison} for massive-massless systems.

In the case of $k_{\mathsmaller{\mathrm{F}}}^{\mathsmaller{m}}/k_{\mathsmaller{\mathrm{F}}}^{\mathsmaller{\mathrm{g}}}\ll1$, we can expand the integrands around $k_{\mathsmaller{\mathrm{F}}}^{\mathsmaller{m}}/k_{\mathsmaller{\mathrm{F}}}^{\mathsmaller{\mathrm{g}}}=0$ and then perform the integration of the lowest order contributions analytically,
\begin{equation}\label{f_ll}
\begin{aligned}
f\left(k_{\mathsmaller{\mathrm{F}}}^{\mathsmaller{m}},k_{\mathsmaller{\mathrm{F}}}^{\mathsmaller{\mathrm{g}}},q_{\mathsmaller{\mathrm{TF}}}\right)=&\frac{1}{\gamma_<^2-1}+\frac{\Theta\left(1-\gamma_<\right)\ln\left(\frac{1+\sqrt{1-\gamma_<^2}}{\gamma_<}\right)}{\left|\gamma_<^2-1\right|^{3/2}}\\
&+\frac{\Theta\left(\gamma_<-1\right)\left[\arctan\left(\frac{1}{\sqrt{\gamma_<^2-1}}\right)-\frac{\pi}{2}\right]}{\left|\gamma_<^2-1\right|^{3/2}}
\end{aligned}
\end{equation}
and
\begin{equation}\label{f1_ll}
\begin{aligned}
f_1&\left(k_{\mathsmaller{\mathrm{F}}}^{\mathsmaller{m}},k_{\mathsmaller{\mathrm{F}}}^{\mathsmaller{\mathrm{g}}},q_{\mathsmaller{\mathrm{TF}}}\right)=\\
&\frac{2\gamma_<^2(4\gamma_<^2-3)}{\sqrt{\left|\gamma_<^2-1\right|}}\Biggl\{\Theta\left(1-\gamma_<\right)\ln\left(\frac{1-\sqrt{1-\gamma_<^2}}{1+\sqrt{1+\gamma_<^2}}\right)\\
&+2\Theta\left(\gamma_<-1\right)\left[\arctan\left(\frac{1}{\sqrt{\gamma_<^2-1}}\right)-\frac{\pi}{2}\right]\Biggr\}\\
&+8\pi\gamma_<^3-16\gamma_<^2-2\pi\gamma_<+\frac{4}{3}
\end{aligned}
\end{equation}
where $\gamma_<=q_{\mathsmaller{\mathrm{TF}}}/(2k_{\mathsmaller{\mathrm{F}}}^{\mathsmaller{m}})$.

Finally, we consider the case of $k_{\mathsmaller{\mathrm{F}}}^{\mathsmaller{m}}/k_{\mathsmaller{\mathrm{F}}}^{\mathsmaller{\mathrm{g}}}\gg1$. Here, we can expand the integrands around $k_{\mathsmaller{\mathrm{F}}}^{\mathsmaller{\mathrm{g}}}/k_{\mathsmaller{\mathrm{F}}}^{\mathsmaller{m}}=0$ and obtain
\begin{equation}\label{f_gg}
\begin{aligned}
f&\left(k_{\mathsmaller{\mathrm{F}}}^{\mathsmaller{m}},k_{\mathsmaller{\mathrm{F}}}^{\mathsmaller{\mathrm{g}}},q_{\mathsmaller{\mathrm{TF}}}\right)=\\
&\frac{2\gamma_<^2-1}{\sqrt{\left|\gamma_>^2-1\right|}}\Biggl\{-\Theta\left(1-\gamma_>\right)\ln\left(\frac{1+\sqrt{1-\gamma_>^2}}{\gamma_>}\right)\\
&+\Theta\left(\gamma_>-1\right)\left[\arctan\left(\frac{1}{\sqrt{\gamma_>^2-1}}\right)-\frac{\pi}{2}\right]\Biggr\}\\
&+\pi\gamma_>-2,
\end{aligned}
\end{equation}
where $\gamma_>=q_{\mathsmaller{\mathrm{TF}}}/(2k_{\mathsmaller{\mathrm{F}}}^{\mathsmaller{\mathrm{g}}})$. Moreover, the main contribution in this limit arises from $f$ because $f\left(k_{\mathsmaller{\mathrm{F}}}^{\mathsmaller{m}},k_{\mathsmaller{\mathrm{F}}}^{\mathsmaller{\mathrm{g}}},q_{\mathsmaller{\mathrm{TF}}}\right)\gg f_1\left(k_{\mathsmaller{\mathrm{F}}}^{\mathsmaller{m}},k_{\mathsmaller{\mathrm{F}}}^{\mathsmaller{\mathrm{g}}},q_{\mathsmaller{\mathrm{TF}}}\right)$.

\bibliographystyle{apsrev}
\bibliography{BibCoulombDrag}

\end{document}